\documentclass[10pt,twocolumn,letterpaper]{article}

\usepackage[pagenumbers]{cvpr} % To force page numbers, e.g. for an arXiv version

\usepackage{graphicx}
\usepackage{amsmath}
\usepackage{amssymb}
\usepackage{booktabs}
\usepackage[accsupp]{axessibility}  % Improves PDF readability for those with disabilities.

\usepackage{tabularx}

\usepackage[pagebackref,breaklinks,colorlinks]{hyperref}
\usepackage{colortbl}  %彩色表格需要加载的宏包
\usepackage{xcolor}
\usepackage{array}   %对表列和表格线的设置需要用到array宏包
\definecolor{lightgray}{gray}{0.95}
\definecolor{color3}{gray}{0.95}
\definecolor{rouse}{rgb}{0.981,0.961,0.941}

\usepackage[capitalize]{cleveref}
\crefname{section}{Sec.}{Secs.}
\Crefname{section}{Section}{Sections}
\Crefname{table}{Table}{Tables}
\crefname{table}{Tab.}{Tabs.}

\def\cvprPaperID{8711} % *** Enter the CVPR Paper ID here
\def\confName{CVPR}
\def\confYear{2023}

\newcommand \footnoteONLYtext[1]
{
	\let \mybackup \thefootnote
	\let \thefootnote \relax
	\footnotetext{#1}
	\let \thefootnote \mybackup
	\let \mybackup \imareallyundefinedcommand
}

\begin{document}

\title{Residual Degradation Learning Unfolding Framework with Mixing Priors across Spectral and Spatial for Compressive Spectral Imaging}

% \and

\author{
  Yubo Dong \quad Dahua Gao $^*$ \quad Tian Qiu \quad Yuyan Li \quad Minxi Yang \quad Guangming Shi \\
  School of Artificial Intelligence, Xidian University \\
  % \texttt{ShawnDong98@stu.xidian.edu.cn}
  {\tt\small \{ybdong, tianqiu, liyuyan, mxyang\}@stu.xidian.edu.cn \quad \{dhgao, gmshi\}@xidian.edu.cn}
}
\maketitle

\footnoteONLYtext{
This work was supported by the National Key Research and Development Program of China (No. 2019YFA0706604),  the Natural Science Foundation (NSF) of China (Nos. 61976169, 62293483). 

* Corresponding author.
}

%%%%%%%%% ABSTRACT
\begin{abstract}
\vspace{-2mm}
% Abstract goes here.
To acquire a snapshot spectral image, coded aperture snapshot spectral imaging (CASSI) is proposed. A core problem of the CASSI system is to recover the reliable and fine underlying 3D spectral cube from the 2D measurement. By alternately solving a data subproblem and a prior subproblem, deep unfolding methods achieve good performance. However, in the data subproblem, the used sensing matrix is ill-suited for the real degradation process due to the device errors caused by phase aberration, distortion; in the prior subproblem,  it is important to design a suitable model to jointly exploit both spatial and spectral priors. In this paper, we propose a Residual Degradation Learning Unfolding Framework (RDLUF), which bridges the gap between the sensing matrix and the degradation process. Moreover, a Mix$S^2$ Transformer is designed via mixing priors across spectral and spatial to strengthen the spectral-spatial representation capability. Finally, plugging the Mix$S^2$ Transformer into the RDLUF leads to an end-to-end trainable neural network RDLUF-Mix$S^2$. Experimental results establish the superior performance of the proposed method over existing ones. Code is available: \url{https://github.com/ShawnDong98/RDLUF_MixS2}
\end{abstract}

%%%%%%%%% BODY TEXT
\vspace*{-6mm}
\section{Introduction}
\vspace*{-2mm}
\label{sec:intro}

\begin{figure}[t]
	\begin{center}
		\begin{tabular}[t]{c} \hspace{-3.8mm} 
			\includegraphics[height=0.3\textwidth, width=0.35\textwidth]{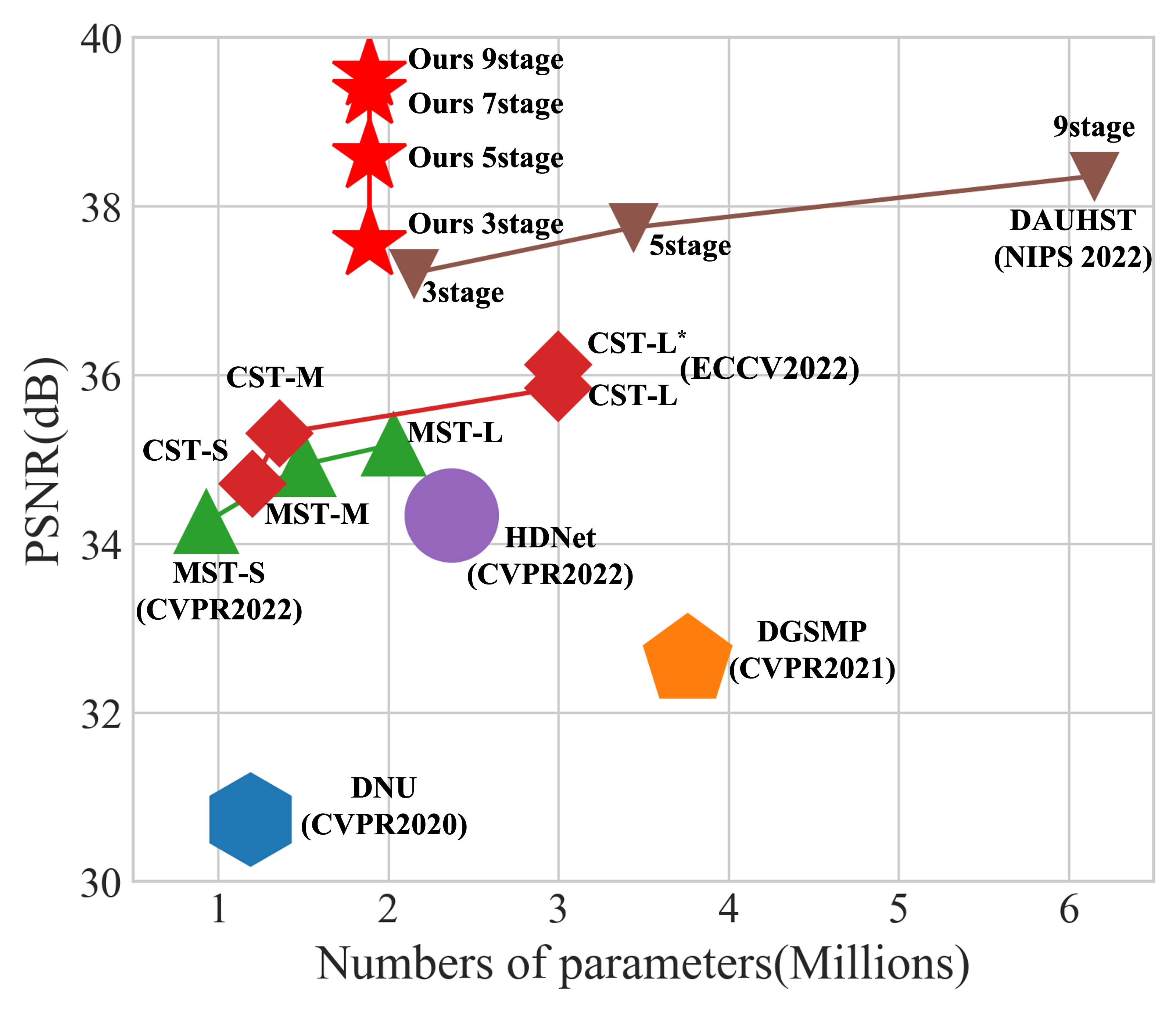}
		\end{tabular}
	\end{center}
	\vspace{-8mm}
	\caption{\small Comparison of PSNR-Parameters with previous HSI reconstruction methods. The PSNR (in dB) is plotted on the vertical axis, while memory cost parameters are represented on the horizontal axis. Our proposed Residual Degradation Learning Unfolding Framework with Mixing priors across Spatial and Spectral (RDLUF-Mix$S^2$) Transformers outperforms previous methods while requiring fewer parameters. }
	\label{fig:teaser}
	\vspace{-7mm}
\end{figure}

With the application of coded aperture snapshot spectral imaging (CASSI) \cite{wagadarikar2008single, arce2013compressive, wang2015high, meng2020end}, it has become feasible to acquire a spectral image using a coded aperture and dispersive elements to modulate the spectral scene. By capturing a multiplexed 2D projection of the 3D data cube, CASSI technique provides an efficient approach for acquiring spectral data. Nonetheless, the reconstruction of an accurate and detailed 3D hyperspectral image (HSI) cube from the 2D measurements poses a fundamental challenge for the CASSI system.

 Based on CASSI, various reconstruction techniques have been developed to reconstruct the 3D HSI cube from 2D measurements. These methods range from model-based techniques \cite{kittle2010multiframe, lin2014spatial, wagadarikar2008single, wang2015dual, yuan2016generalized, liu2018rank, wang2016adaptive, zhang2019computational}, to end-to-end approaches \cite{miao2019net, meng2020end, hu2022hdnet, cai2022mask, lin2022coarse}, and deep unfolding methods \cite{wang2019hyperspectral, wang2020dnu, huang2021deep}. Among them,  deep unfolding methods have demonstrated superior performance by transferring conventional iterative optimization algorithms into a series of deep neural network (DNN) blocks. Typically, the deep unfolding methods tackle a data subproblem and a prior subproblem iteratively.

The data subproblem is highly related to the degradation process. The ways to acquire the degradation matrix in the data subproblem can be classified into two types, the first directly uses the sensing matrix as the degradation matrix \cite{ma2019deep, meng2020gap, wang2020dnu} and the other learns the degradation matrix using a neural network \cite{ huang2021deep, mou2022deep, zhang2022herosnet}. However, since the sensing matrix is obtained from the equidistant lasers of different wavelengths on the sensor, it cannot reflect the device errors caused by phase aberration, distortion and alignment of the continuous spectrum. Thus, the earlier kind does not take into account the gap between the sensing matrix and the degradation process. In the latter type, directly modeling the degradation process is challenging. Considering the challenge of optimizing the original, unreferenced mapping, it is preferable to focus on optimizing the residual mapping. Therefore, we explicitly model the degradation process as residual learning with reference to the sensing matrix.

For the prior subproblem, a denoiser is trained to represent the regularization term as a denoising problem in an implicit manner, typically implemented as an end-to-end neural network.  Recently, Spectral-wise Multi-head Self-Attention (S-MSA) has been introduced to model long-range dependency in the spectral dimension. However, S-MSA may neglect spatial information that is crucial for generating high-quality HSI images, due to its implicit modeling of spatial dependency. To this end, the integration of Convolutional Neural Networks (CNNs) with S-MSA can provide an ideal solution as CNNs have the inductive bias of modeling local similarity, thus enhancing the spatial modeling capabilities of S-MSA. To achieve this, we propose a multiscale convolution branch that processes visual information at multiple scales and then aggregates it to enable simultaneous feature abstraction from different scales, thereby capturing more textures and details.

In this paper, we first unfold the Proximal Gradient Descent (PGD) algorithm under the framework of maximum a posteriori theory for HSI reconstruction. Then, we integrate the residual degradation learning strategy into the data subproblem of PGD, which briges the gap between the sensing matrix and the degradation process, leading to our Residual Degradation Learning Unfolding Framework (RDLUF). Secondly, a multiscale convolution called Lightweight Inception is combined with spectral self-attention in a parallel design to address the problem of weak spatial modeling ability of S-MSA. To provide complementary clues in the spectral and spatial branches, we propose a bi-directional interaction across branches, which enhance the modeling ability in spectral and spatial dimensions respectively, resulting in our Mixing priors across Spatial and Spectral (Mix$S^2$) Transformer. Finally, plugging the Mix$S^2$ Transformer into the RDLUF as the denoiser of the prior subproblem leads to an end-to-end trainable neural network RDLUF-Mix$S^2$. Equipped with the proposed techniques, RDLUF-Mix$S^2$ achieves state-of-the-art (SOTA) performance on HSI reconstruction, as shown in Fig. \ref{fig:teaser}.

% Update the cvpr.cls to do the following automatically.
% For this citation style, keep multiple citations in numerical (not
% chronological) order, so prefer \cite{Alpher03,Alpher02,Authors14} to
% \cite{Alpher02,Alpher03,Authors14}.

%------------------------------------------------------------------------
\vspace{-3mm}
\section{Related Work}
\label{sec:related}
\vspace{-2mm}
\subsection{Deep Unfolding HSI Reconstruction}

 Generally, when attempting to reconstruct HSI, model-based techniques \cite{wagadarikar2008single, kittle2010multiframe, liu2018rank, wang2016adaptive, zhang2019computational, yuan2016generalized, tan2015compressive, arce2013compressive, figureido2007gradient} adopt a Bayesian perspective and cast it as an optimization problem of maximizing the posterior probability (MAP). The optimization algorithms commonly used are HQS \cite{he2013half}, ADMM \cite{boyd2011distributed}, and PGD \cite{beck2009fast}. 
Typically, such techniques disentangle the data fidelity and the regularization terms in the objective function, leading to an iterative procedure that alternates between solving a data subproblem and a prior subproblem.

The main idea of deep unfolding methods is that model-based iterative optimization algorithms can be implemented equivalently by a stack of recurrent DNN blocks. Such design was originally applied in deep plug-and-play methods \cite{meinhardt2017learning,ryu2019plug,yuan2020plug,zhang2017learning,zhang2019deep}, which utilize a trained denoiser to implicitly express the prior subproblem as a denoising problem. Inspired by plug-and-play, deep unfolding methods are trained end-to-end by jointly optimizing trainable denoisers for specific tasks.   GAP-net \cite{meng2020gap} unfolds the generalized alternating projection algorithm and employs trained auto-encoder-based denoisers. DGSMP \cite{huang2021deep} introduces an unfolding model estimation framework, which leverages the learned gaussian scale mixture prior to improve model performance. These methods typically employ the sensing matrix as the degradation matrix or a neural network is utilized to learn the degradation matrix in the data subproblem.

\vspace{-2mm}
\subsection{Methods for Exploiting Spectral and Spatial Priors}
\vspace{-2mm}

 Model-based approaches often utilize manually crafted priors such as the total variation prior \cite{yuan2016generalized}. On the other hand, sparse-based methods \cite{kittle2010multiframe, lin2014spatial, wagadarikar2008single} rely on the assumption that HSIs exhibit sparse representations and use $\ell_1$ sparsity to regularize the solution. In addition to that, non-local based techniques \cite{liu2018rank, wang2016adaptive, zhang2019computational} take into account the strong long-range dependency among HSI pixels to achieve more accurate results.

Previous researchers have utilized end-to-end neural networks to leverage data-driven priors, as demonstrated in \cite{wang2019hyperspectral, wang2020dnu}. CNN-based methods exhibit powerful local similarity modeling capabilities. $\lambda$-net \cite{miao2019net} reconstructs the HSIs via a two-step process. In \cite{meng2020end},  TSA-Net is proposed to exploit spatial-spectral correlation. Despite their effectiveness in some tasks, CNN-based techniques may have limitations in identifying non-local similarities as a result of their inductive biases.  To address these shortcomings, transformer-based methods have been proposed, such as \cite{cai2022mask, lin2022coarse}. These methods utilize multi-head self-attention mechanisms to model the long-range spatial and spectral dependency in HSIs. For instance, S-MSA in MST \cite{cai2022mask} computes dependency across spectral to generate an attention map that encodes global context implicitly. However, this approach may lead to a loss of significant spatial information related to textures and structures, which is crucial in generating high-quality HSI images.

%------------------------------------------------------------------------
\vspace{-3mm}
\section{Method}
\label{sec:method}
\vspace{-2mm}
\subsection{Problem Formulation}
\vspace{-2mm}

% The main components of the CASSI system \cite{arce2013compressive, wagadarikar2008single} that contribute to the degradation process include the physical mask, the dispersive prism and the 2D imaging sensor. The mask $M \in \mathbb{R}^{H \times W}$ is used to modulate HSI signals $X \in \mathbb{R}^{H \times W \times N_{\lambda}}$. Therefore, the $n_{\lambda}^{th}$ wavelength of the modulated image can be represented as:

The degradation of CASSI system \cite{arce2013compressive, wagadarikar2008single} can be attributed to various factors, including the physical mask, dispersive prism, and 2D imaging sensor. The physical mask, denoted by $M \in \mathbb{R}^{H \times W}$, acts as a modulator for the HSI signal $X \in \mathbb{R}^{H \times W \times N_{\lambda}}$, thereby enabling the representation of the $n_{\lambda}^{th}$ wavelength of the modulated image:

\begin{equation}
    X_{n_\lambda}' = M \odot X_{n_{\lambda}},
    \label{eq mask}
\end{equation}
where $\odot$ represents the element-wise product. Consequently, the modulated HSI $X'$ are shifted during the dispersion process, which can be expressed as:

\vspace{-2mm}

\begin{equation}
    X''(h, w, n_{\lambda}) = X'(h, w+d_{n_\lambda}, n_{\lambda}),
    \label{eq dispersion process}
\end{equation}
where $X'' \in \mathbb{R}^{H \times (W + d_{N_\lambda}) \times N_\lambda}$,   $d_{n_{\lambda}}$ represents the shifted distance of the $n_{\lambda}^{th}$ wavelength. At last, the imaging sensor captures the shifted image into a 2D measurement. This process can be formulated as follow:

\vspace{-2mm}

\begin{equation}
    Y = \sum_{n_{\lambda}=1}^{N_{\lambda}} X''_{n_\lambda},
    \label{eq sum process}
\end{equation}
the 3D HSI cube is degraded to 2D measurement $Y \in \mathbb{R}^{H \times (W + d_{N_\lambda})}$ after the sum operator, and the spatial dimensions increased as the dispersion process. As such, considering the measurement noise, the matrix-vector form of Eq. \ref{eq sum process} can be formulated as:

\vspace{-2mm}

\begin{equation}
    y = \Phi x + n,
    \label{eq matrix-vector form}
\end{equation}
where $x$ is the original HSI, y is the degraded measurement, $\Phi$ is the sensing matrix, generally consider it as all the degraded operators (Eq. (\ref{eq mask}, \ref{eq dispersion process}, \ref{eq sum process})), and $n$ represents the additive noise. HSI restoration aims to recover the high-quality image $x$ from its degraded measurement $y$, which is typically an ill-posed problem.

Model-based methods (e.g.,\cite{wagadarikar2008single, kittle2010multiframe, liu2018rank, wang2016adaptive, zhang2019computational, yuan2016generalized, tan2015compressive, arce2013compressive, figureido2007gradient}) usually formulate HSI reconstruction as a Bayesian problem, solving Eq. (\ref{eq matrix-vector form}) under a unified MAP framework:

\vspace{-2mm}

\begin{equation}
    \hat x = \mathop{\text{argmax}}_{x} \log P(x \mid y) = \mathop{\text{argmax}}_{x} \log P(y \mid x) + \log P(x),
    \label{eq MAP}
\end{equation}
where $\log P(y \mid x)$ and $\log P(x)$ represent the data fidelity and the regularization term, respectively. The data fidelity term is usually defined as an $\ell_2$ norm, expressing Eq. (\ref{eq MAP}) as the following energy function:

\vspace{-2mm}

\begin{equation}
    \hat x = \mathop{\text{argmin}}_x \frac{1}{2} \|y - \Phi x \|_2^2 + \lambda J(x). 
    \label{eq energy function}
\end{equation}

The PGD algorithm approximatively expresses Eq. (\ref{eq  energy function}) as an iterative convergence problem through the following iterative function:

\vspace{-2mm}

\begin{equation}
    \color{cyan}\hat x^k =  \mathop{\text{argmin}}_x \frac{1}{2 \rho} \| x - {\color{red}(\hat x^{k-1} - \rho  \Phi^T (\Phi \hat x^{k-1} - y))} \|_2^2 + \lambda J(x), 
    \label{eq PGD}
\end{equation}
where $\hat x^k$ refers to the output of the $k$-th iteration,  $\rho$ is the step size. Mathematically, the {\color{red}red} part of the above function is a gradient descent operation and the {\color{cyan}blue} part can be solved by the proximal operator $\text{prox}_{\lambda,J}$. Thus, it leads to a data subproblem and a prior subproblem, i.e., gradient descent (Eq. (\ref{eq gd})) and proximal mapping (Eq. (\ref{eq pm})):

\vspace{-2mm}

\begin{subequations}
    \begin{gather}
        {\color{red} v^{k} = \hat x^{k-1} - \rho \Phi^T (\Phi \hat x^{k-1} - y),}
        \label{eq gd} \\
        {\color{cyan} \hat x^k = \text{prox}_{\lambda, J}(v^k).}
        \label{eq pm}
    \end{gather}
\end{subequations}

The PGD algorithm iteratively updates $v^k$ and $\hat x^k$ until convergence. The model-based algorithms mainly suffer from two issues in the CASSI system. Firstly, since the sensing matrix is obtained from the equidistant lasers of different wavelengths on the sensor, it cannot reflect the device errors caused by phase aberration, distortion and alignment of the continuous spectrum. Therefore, there exist the gap between the sensing matrix and the degradation matrix that used in the data subproblem. Secondly, the handcrafted priors have to tweak parameters manually, resulting in limited representation abilities in addition to the slow reconstruction speed. To address these issues, we unfold the PGD algorithm by DNNs and integrate residual degradation learning into the gradient descent step.

\begin{figure*}[tp]
    \centering
    \includegraphics[width=\linewidth]{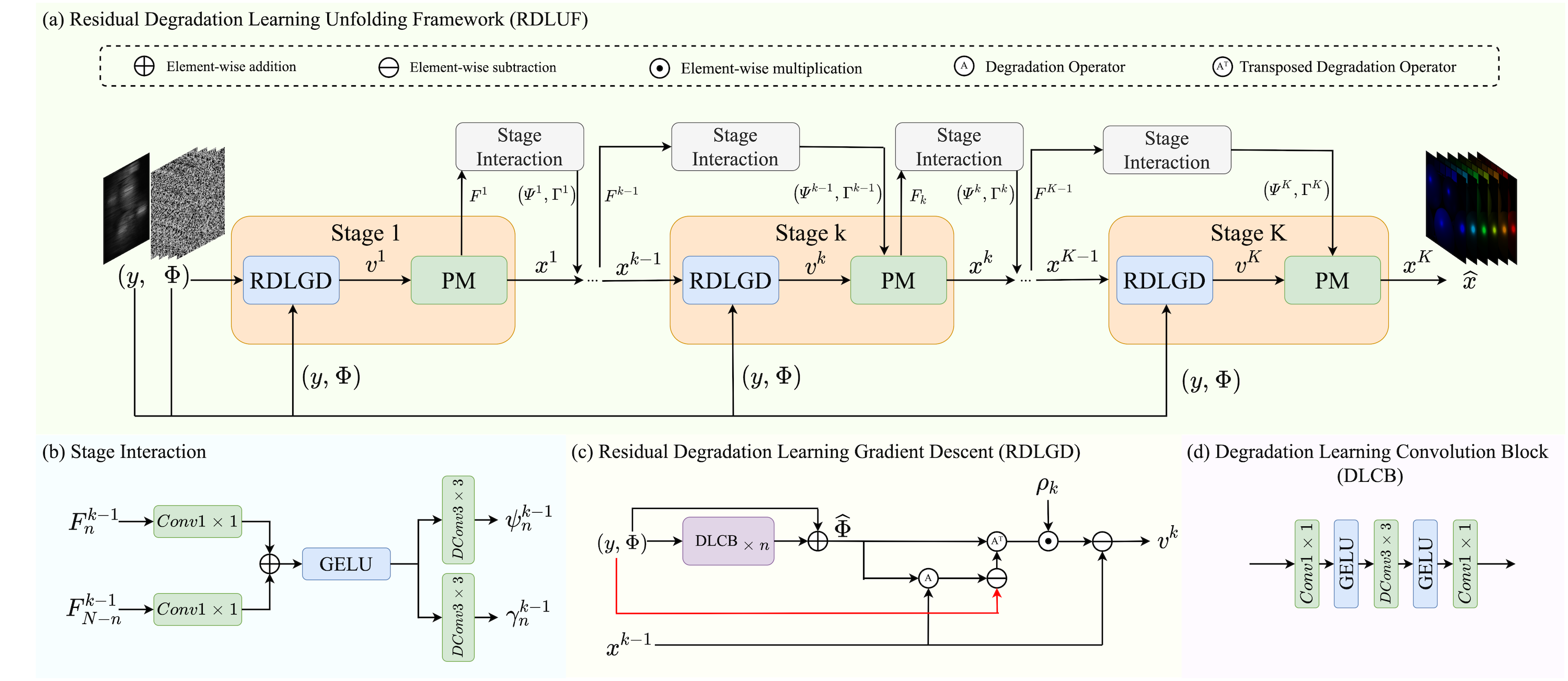}
    \vspace*{-5mm}
    \caption{\small Our proposed RDLUF comprises $K$ stages (iterations). $DConv$ denotes the depth-wise convolution. RDLGD calculates the degradation matrix $\hat \Phi$ by utilizing compressed data $y$ and sensing matrix $\Phi$, and executes the gradient descent with a learnable step size parameter $\rho_k$. There is a stage interaction between stages.} 
    \label{fig:dluf}
    \vspace*{-5mm}
\end{figure*}

\vspace{-2mm}
\subsection{Residual Degradation Learning Unfolding Framework}
\vspace{-2mm}

The whole architecture of the proposed RDLUF is presented in Fig. \ref{fig:dluf} (a), which is an unfolding framework of the PGD algorithm based on DNNs. Our RDLUF is composed of several repeated stages. Each stage contains a Residual Degradation Learning Gradient Descent (RDLGD) module and a Proximal Mapping (PM) module, corresponding to the gradient descent (Eq. (\ref{eq gd})) and the proximal mapping (Eq. (\ref{eq pm})) in an iteration step of the PGD algorithm, respectively. Additionally, there is a stage interaction between two stages to rich features and stable optimization in a spatial adaptive normalization manner.

\textbf{Residual Degradation Learning  Gradient Descent.} In a snapshot compressive imaging system, since the sensing matrix is obtained from the equidistant lasers of different wavelengths on the sensor, it cannot reflect the device errors caused by phase aberration, distortion and alignment of the continuous spectrum. Therefore, there exists the gap between the sensing matrix $\Phi$ and the degradation matrix $\hat \Phi$.  Previous methods proposed learning the degradation matrix using a neural network. However, it is challenging to directly model the degradation process. It is easier to optimize the residual mapping than to optimize the original, unreferenced mapping. Therefore, instead of directly learning the degradation matrix, we propose the RDLGD module to estimate the residual between the sensing matrix $\Phi$ and the degradation matrix $\hat \Phi$  from the compressed measurement $y$ and the sensing matrix $\Phi$. The RDLGD's architecture is shown in Fig. \ref{fig:dluf} (c), which is consist of several Degradation Learning Convolution Blocks (DLCBs). The DLCB is illustrated in Fig. \ref{fig:dluf} (d). The degradation matrix can be calculated as $\hat \Phi = \Phi + \mathcal{R}(y, \Phi)$, where $\mathcal{R}$ represents the several cascaded DLCB, $y$ is the compressed measurement, $\Phi$ is the sensing matrix.

Therefore, the gradient descent step in our proposed RDLUF can be expressed as:

\vspace{-2mm}

\begin{equation}
    \color{red} v^k = \hat x^{k-1} - \rho^k {{\hat \Phi}^{\top^k}}({\hat \Phi}^k \hat x^{k-1} - y),
    \label{eq hat gd}
\end{equation}
where $\rho$ is a learnable parameter, $k$ represents the number of stage.

\textbf{Proximal Mapping.} For the solution of Eq. (\ref{eq pm}), it is known that, from a Bayesian perspective, it corresponds to a denoising problem \cite{chan2016plug, yuan2020plug}. In this context, we have designed the Mix$S^2$ Transformer as the PM module, which effectively mixes priors across spectral and spatial, as presented in Fig. \ref{fig:MixS2 Transformer} (a). More details will be illustrated in the next section.

\textbf{Stage Interaction.} To reduce the loss of information,  enrich the features of each stage and ease the network optimization procedure, we propose the stage interaction module, which generates modulation parameters from the previous stage features to normalize the current stage features in a spatial adaptive normalization manner \cite{park2019semantic, mou2022deep}. More details can be found in Appendix.

\vspace{-2mm}
\subsection{Mixing priors across Spectral and Spatial Transformer}
\vspace{-2mm}

In this section, we explain the proposed model Mix$S^2$ Transformer in detail.

\begin{figure*}[tp]
    \centering
    \includegraphics[width=\linewidth, height=.7\linewidth]{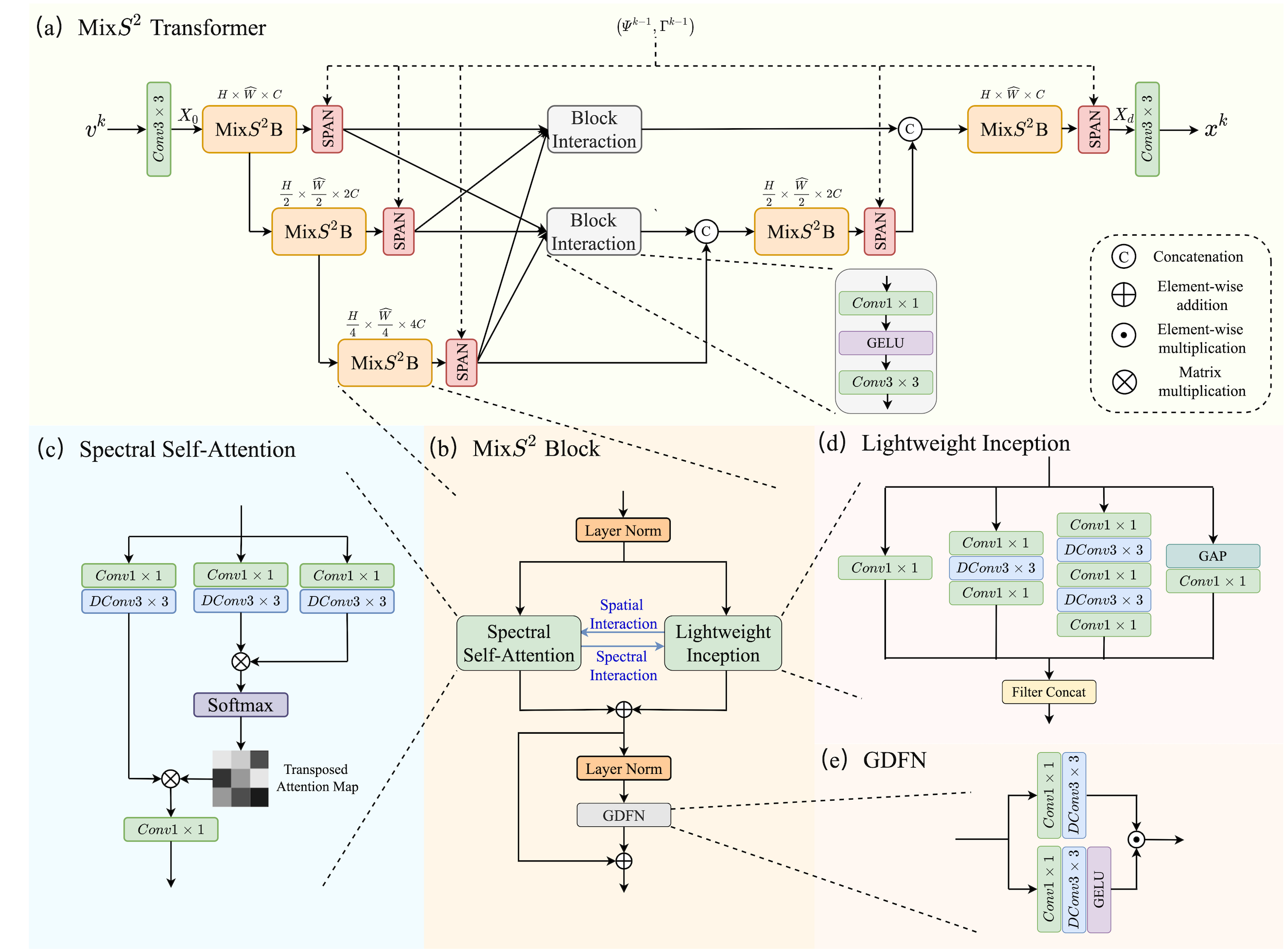}
    \vspace*{-7mm}
    \caption{\small Diagram of the Mix$S^2$ Transformer. (a) Mix$S^2$ Transformer adopts a U-shaped structure with block interactions. (b) The basic unit of the Mix$S^2$ Transformer, Mix$S^2$ block. (c) The structure of the spectral self-attention branch. (d) The structure of the lightweight inception branch. (e) The components of the gated-Dconv feed-forward network (GDFN).}. 
    \label{fig:MixS2 Transformer}
    \vspace*{-7mm}
\end{figure*}

\textbf{Network Architecture.} As shown in Fig. \ref{fig:MixS2 Transformer} (a), the Mix$S^2$ Transformer adopts a U-shaped structure consists of several basic unit Mix$S^2$ blocks. There are up- and down-sampling modules between Mix$S^2$ blocks. We introduce the block interaction to reduce the loss of information caused by up- and down-sampling operations. Different scale features are interpolated to the same scale in block interaction. Firstly, the Mix$S^2$ Transformer uses a $Conv3 \times 3$ to map $v_k$  into shallow features $X_0 \in \mathbb{R}^{H \times \hat W \times C}$, where $\hat W = W + d_{N_\lambda}$. Secondly, $X_0$ passes through all the Mix$S^2$ blocks and block interactions to be embedded into deep features $X_d \in \mathbb{R}^{H \times \hat W \times C}$.  Finally, a $Conv3 \times 3$ operates on $X_d$ to generate the denoised image $x_k$. 

\textbf{Mixing priors across Spectral and Spatial Block.} The most important component of the Mix$S^2$ Transformer is the Mix$S^2$ block as shown in Fig. \ref{fig:MixS2 Transformer} (b), which consists of two layer normalization, a spectral self-attention branch and a lightweight inception branch in a parallel design with a bi-directional interaction, and a gated-Dconv feed-forward network \cite{zamir2022restormer} that is detailed in Fig. \ref{fig:MixS2 Transformer} (e). The up- and down-sampling modules are both a $Conv 3 \times 3$ and a bilinear interpolation with different scale factors. The details of the spectral self-attention branch, the lightweight inception branch, and the bi-directional interaction are described following. 

\textbf{Spectral Self-Attention Branch.} The spectral self-attention branch is following \cite{zamir2022restormer}, which $query$($\mathrm{Q}$), $key$($\mathrm{K}$) and $value$($\mathrm{V}$) are embedded with a $Conv 1 \times 1$ and a $DConv 3 \times 3$ different from \cite{cai2022mask}. The key component of the spectral self-attention branch is S-MSA. Fig. \ref{fig:MixS2 Transformer} (c) shows the spectral self-attention branch. The input $\mathrm{X}_{in} \in \mathbb{R}^{H \times W \times C}$ is first embedded, yielding $\mathrm{Q}=W_d^Q W_p^Q \mathrm{X}$, $\mathrm{K} = W_d^K W_p^K \mathrm{X}$ and $\mathrm{V} = W_d^V W_p^V \mathrm{X}$. Where $W_p^{(\cdot)}$ is the $Conv1 \times 1$ and $W_d^{(\cdot)}$ is the $DConv3 \times 3$. Next, the S-MSA reshapes the query and key projections such that their dot-product interaction generates a transposed attention map $\mathrm{A}$ of size $\mathbb{R}^{C \times C}$. Overall, the S-MSA process is defined as:

\vspace{-2mm}

\begin{equation}
    \begin{aligned}
        & \mathrm{X} = W_p \text{Attention}(\mathrm{Q}, \mathrm{K}, \mathrm{V}), \\
        &\text{Attention}(\mathrm{Q}, \mathrm{K}, \mathrm{V}) = \mathrm{V} \text{Softmax}(\mathrm{K} \mathrm{Q} / \alpha),
    \end{aligned}
\end{equation}
where $\mathrm{Q} \in \mathbb{R}^{HW \times C}$; $\mathrm{K} \in \mathbb{R}^{C \times HW}$; and $\mathrm{V} \in \mathbb{R}^{HW \times C}$ matrices are obtained after reshaping tensors from the original size $\mathbb{R}^{H \times W \times C}$. Here, $\alpha$ is a learnable scaling parameter to control the magnitude of the dot product $\mathrm{K}$ and $\mathrm{Q}$ before applying the softmax function.

\textbf{Lightweight Inception Branch.} Corresponding to the multiscale convolution, the structure of the lightweight inception branch follows \cite{szegedy2016rethinking, szegedy2017inception}, but the difference is that we use $DConv 3 \times 3$ instead of $Conv 3 \times 3$, as illustrated in Fig. \ref{fig:MixS2 Transformer} (d).  The visual information is processed at various scales and then aggregated so that the next layer can abstract features from different scales simultaneously and hence capture more textures and details.

\begin{figure}[tp]
    \centering
    \includegraphics[width=\linewidth]{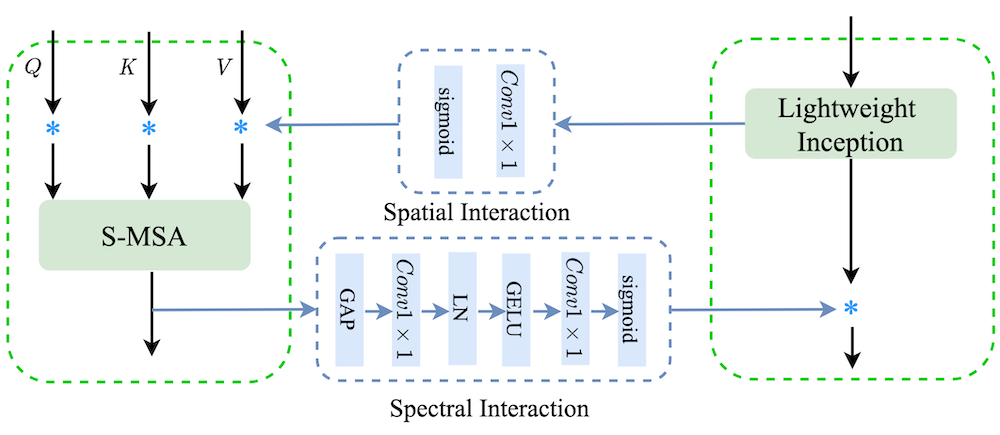}
    \vspace*{-7mm}
    \caption{\small \textbf{Detailed design of the bi-directional interaction}. The spatial/spectral interaction provides spatial/spectral context extracted by Spectral Self-Attention/Lightweight Inception to the other path.}. 
    \vspace*{-10mm}
    \label{fig:BI}
\end{figure}

\textbf{Bi-directional Interaction.} Familiar with \cite{chen2022mixformer}, we introduce the bi-directional interaction across spectral and spatial branches. For the spatial interaction, the output features of the lightweight inception module pass through a $Conv 1 \times 1$ with a sigmoid activation to get a spatial attention map,  applying to the $query(\mathrm{Q})$, $key(\mathrm{K})$ and $value(\mathrm{V})$ in a spatial attention manner. For the spectral interaction, the output features of the spectral self-attention module are fed into a spectral attention module\cite{hu2018squeeze} to obtain a spectral attention weight, applying to the output features of the lightweight inception module in a spectral attention manner. More details are shown in Fig. \ref{fig:BI}.

% \textbf{Block Interaction.} The encoder-decoder networks \cite{brooks2019unprocessing, chen2018learning, kupyn2019deblurgan, ronneberger2015u} first gradually map the input to low-resolution representations, and then progressively apply reverse mapping to recover the original resolution.  While these models effectively encode multi-scale information, they are prone to sacrificing spatial details due to the repeated use of downsampling operations. To address this issue, inspired by \cite{cho2021rethinking}, we introduce the block interaction to allow information flow from different scales within a single Mix$S^2$ Transformer. Each block interaction takes the output of all encoder blocks as an input and combines multiscale features using convolutional layers. The output of the block interaction is delivered to its corresponding decoder block.

%------------------------------------------------------------------------
\vspace{-3mm}
\section{Experiments}
\label{sec:experiments}

\begin{table*}[t]
	\renewcommand{\arraystretch}{1.0}
	\newcommand{\tabincell}[2]{\begin{tabular}{@{}#1@{}}#2\end{tabular}}
	% \caption{SSIM values by different algorithms on 10 synthetic data.}
	\centering
	\resizebox{0.99\textwidth}{!}
	{
		\centering
		% \begin{tabular}{c|c|c|c|c|c|c|>{\columncolor{lightgray}}c}
		\begin{tabular}{cccccccccccc}
			\toprule[0.2em]
                \rowcolor{lightgray}
			~~~~~Algorithms~~~~~
			& ~~~~~Scene1~~~~~
			& ~~~~~Scene2~~~~~
			& ~~~~~Scene3~~~~~
			& ~~~~~Scene4~~~~~
			& ~~~~~Scene5~~~~~
			& ~~~~~Scene6~~~~~
			& ~~~~~Scene7~~~~~
			& ~~~~~Scene8~~~~~
			& ~~~~~Scene9~~~~~
			& ~~~~~Scene10~~~~~
			& ~~~~Avg~~~~
			\\
			\midrule
			TwIST \cite{bioucas2007new}
			&\tabincell{c}{25.16\\0.700}
			&\tabincell{c}{23.02\\0.604}
			&\tabincell{c}{21.40\\0.711}
			&\tabincell{c}{30.19\\0.851}
			&\tabincell{c}{21.41\\0.635}
			&\tabincell{c}{20.95\\0.644}
			&\tabincell{c}{22.20\\0.643}
			&\tabincell{c}{21.82\\0.650}
			&\tabincell{c}{22.42\\0.690}
			&\tabincell{c}{22.67\\0.569}
			&\tabincell{c}{23.12\\0.669}
			\\
			\midrule
			GAP-TV \cite{yuan2016generalized}
			&\tabincell{c}{26.82\\0.754}
			&\tabincell{c}{22.89\\0.610}
			&\tabincell{c}{26.31\\0.802}
			&\tabincell{c}{30.65\\0.852}
			&\tabincell{c}{23.64\\0.703}
			&\tabincell{c}{21.85\\0.663}
			&\tabincell{c}{23.76\\0.688}
			&\tabincell{c}{21.98\\0.655}
			&\tabincell{c}{22.63\\0.682}
			&\tabincell{c}{23.10\\0.584}
			&\tabincell{c}{24.36\\0.669}
			\\
			\midrule
			DeSCI \cite{liu2018rank}
			&\tabincell{c}{27.13\\0.748}
			&\tabincell{c}{23.04\\0.620}
			&\tabincell{c}{26.62\\0.818}
			&\tabincell{c}{34.96\\0.897}
			&\tabincell{c}{23.94\\0.706}
			&\tabincell{c}{22.38\\0.683}
			&\tabincell{c}{24.45\\0.743}
			&\tabincell{c}{22.03\\0.673}
			&\tabincell{c}{24.56\\0.732}
			&\tabincell{c}{23.59\\0.587}
			&\tabincell{c}{25.27\\0.721}
			\\
			\midrule
			HSSP \cite{wang2019hyperspectral}
			&\tabincell{c}{31.48\\0.858}
			&\tabincell{c}{31.09\\0.842}
			&\tabincell{c}{28.96\\0.823}
			&\tabincell{c}{34.56\\0.902}
			&\tabincell{c}{28.53\\0.808}
			&\tabincell{c}{30.83\\0.877}
			&\tabincell{c}{28.71\\0.824}
			&\tabincell{c}{30.09\\0.881}
			&\tabincell{c}{30.43\\0.868}
			&\tabincell{c}{28.78\\0.842}
			&\tabincell{c}{30.35\\0.852}
			\\
			\midrule
			DNU \cite{wang2020dnu}
			&\tabincell{c}{31.72\\0.863}
			&\tabincell{c}{31.13\\0.846}
			&\tabincell{c}{29.99\\0.845}
			&\tabincell{c}{35.34\\0.908}
			&\tabincell{c}{29.03\\0.833}
			&\tabincell{c}{30.87\\0.887}
			&\tabincell{c}{28.99\\0.839}
			&\tabincell{c}{30.13\\0.885}
			&\tabincell{c}{31.03\\0.876}
			&\tabincell{c}{29.14\\0.849}
			&\tabincell{c}{30.74\\0.863}
			\\
			\midrule
			DGSMP \cite{huang2021deep}
			&\tabincell{c}{33.26\\0.915}
			&\tabincell{c}{32.09\\0.898}
			&\tabincell{c}{33.06\\0.925}
			&\tabincell{c}{40.54\\0.964}
			&\tabincell{c}{28.86\\0.882}
			&\tabincell{c}{33.08\\0.937}
			&\tabincell{c}{30.74\\0.886}
			&\tabincell{c}{31.55\\0.923}
			&\tabincell{c}{31.66\\0.911}
			&\tabincell{c}{31.44\\0.925}
			&\tabincell{c}{32.63\\0.917}
			\\
			\midrule
			HDNet \cite{hu2022hdnet}
			&\tabincell{c}{35.14\\0.935}
			&\tabincell{c}{35.67\\0.940}
			&\tabincell{c}{36.03\\0.943}
			&\tabincell{c}{42.30\\0.969}
			&\tabincell{c}{32.69\\0.946}
			&\tabincell{c}{34.46\\0.952}
			&\tabincell{c}{33.67\\0.926}
			&\tabincell{c}{32.48\\0.941}
			&\tabincell{c}{34.89\\0.942}
			&\tabincell{c}{32.38\\0.937}
			&\tabincell{c}{34.97\\0.943}
			\\
			\midrule
			MST-L \cite{cai2022mask}
			&\tabincell{c}{35.40\\0.941}
			&\tabincell{c}{35.87\\0.944}
			&\tabincell{c}{36.51\\0.953}
			&\tabincell{c}{42.27\\0.973}
			&\tabincell{c}{32.77\\0.947}
			&\tabincell{c}{34.80\\0.955}
			&\tabincell{c}{33.66\\0.925}
			&\tabincell{c}{32.67\\0.948}
			&\tabincell{c}{35.39\\0.949}
			&\tabincell{c}{32.50\\0.941}
			&\tabincell{c}{35.18\\0.948}
			\\
			\midrule
			CST-L$^*$ \cite{lin2022coarse}
			&\tabincell{c}{35.96\\0.949}
			&\tabincell{c}{36.84\\0.955}
			&\tabincell{c}{38.16\\0.962}
			&\tabincell{c}{42.44\\0.975}
			&\tabincell{c}{33.25\\0.955}
			&\tabincell{c}{35.72\\0.963}
			&\tabincell{c}{34.86\\0.944}
			&\tabincell{c}{34.34\\0.961}
			&\tabincell{c}{36.51\\0.957}
			&\tabincell{c}{33.09\\0.945}
			&\tabincell{c}{36.12\\0.957}
			\\
			\midrule
			DAUHST-9stg \cite{cai2022degradation}
			&\tabincell{c}{37.25\\0.958}
			&\tabincell{c}{39.02\\0.967}
			&\tabincell{c}{41.05\\0.971}
			&\tabincell{c}{46.15\\0.983 }
			&\tabincell{c}{35.80\\0.969}
			&\tabincell{c}{37.08\\0.970}
			&\tabincell{c}{37.57\\0.963}
			&\tabincell{c}{35.10\\0.966}
			&\tabincell{c}{40.02\\0.970}
			&\tabincell{c}{34.59\\0.956}
			&\tabincell{c}{38.36\\0.967}
			\\
                \midrule
                \rowcolor{rouse}
			\bf{Ours 3stage}
                &\tabincell{c}{36.67\\0.953}
			&\tabincell{c}{38.48\\0.965}
			&\tabincell{c}{40.63\\0.971}
			&\tabincell{c}{46.04\\0.986}
			&\tabincell{c}{34.63\\0.963}
			&\tabincell{c}{36.18\\0.966}
			&\tabincell{c}{35.85\\0.951}
			&\tabincell{c}{34.37\\0.963}
			&\tabincell{c}{38.98\\0.966}
			&\tabincell{c}{33.73\\0.950}
			&\tabincell{c}{37.56\\0.963}
                \\
                \midrule
                \rowcolor{rouse}
                \bf{Ours 5stage}
                &\tabincell{c}{37.30\\0.960}
			&\tabincell{c}{39.39\\0.971}
			&\tabincell{c}{42.06\\0.975}
			&\tabincell{c}{46.89\\0.988}
			&\tabincell{c}{35.74\\0.969}
			&\tabincell{c}{37.03\\0.971}
			&\tabincell{c}{37.05\\0.959}
			&\tabincell{c}{35.18\\0.968}
			&\tabincell{c}{40.64\\0.973}
			&\tabincell{c}{34.58\\0.957}
			&\tabincell{c}{38.59\\0.969}
                \\
                \midrule
                \rowcolor{rouse}
                \bf{Ours 7stage}
                &\tabincell{c}{37.65\\0.963}
			&\tabincell{c}{40.45\\0.976}
			&\tabincell{c}{43.00\\0.978}
			&\tabincell{c}{47.40\\0.990}
			&\tabincell{c}{36.78\\0.974}
			&\tabincell{c}{\bf{37.56}\\0.974}
			&\tabincell{c}{38.25\\0.967}
			&\tabincell{c}{\bf{35.86}\\\bf{0.971}}
			&\tabincell{c}{41.71\\0.978}
			&\tabincell{c}{34.83\\0.959}
			&\tabincell{c}{39.35\\0.973}
                \\
                \midrule
                \rowcolor{rouse}
                \bf{Ours 9stage}
                &\tabincell{c}{\bf{37.94}\\\bf{0.966}}
			&\tabincell{c}{\bf{40.95}\\\bf{0.977}}
			&\tabincell{c}{\bf{43.25}\\\bf{0.979}}
			&\tabincell{c}{\bf{47.83}\\\bf{0.990}}
			&\tabincell{c}{\bf{37.11}\\\bf{0.976}}
			&\tabincell{c}{37.47\\\bf{0.975}}
			&\tabincell{c}{\bf{38.58}\\\bf{0.969}}
			&\tabincell{c}{35.50\\0.970}
			&\tabincell{c}{\bf{41.83}\\\bf{0.978}}
			&\tabincell{c}{\bf{35.23}\\\bf{0.962}}
			&\tabincell{c}{\bf{39.57}\\\bf{0.974}}
                \\
			\bottomrule[0.2em]
		\end{tabular}
	}
	\vspace{-1mm}
	\caption{The PSNR (upper entry in each cell) in dB and SSIM (lower entry in each cell) results of the test methods on 10 scenes. RDLUF-Mix$S^2$ significantly surpasses other competitors.}
	\vspace{-3mm}
	\label{Tab:performance}
\end{table*}

\begin{figure*}[ht]
    \centering
    \includegraphics[width=\linewidth]{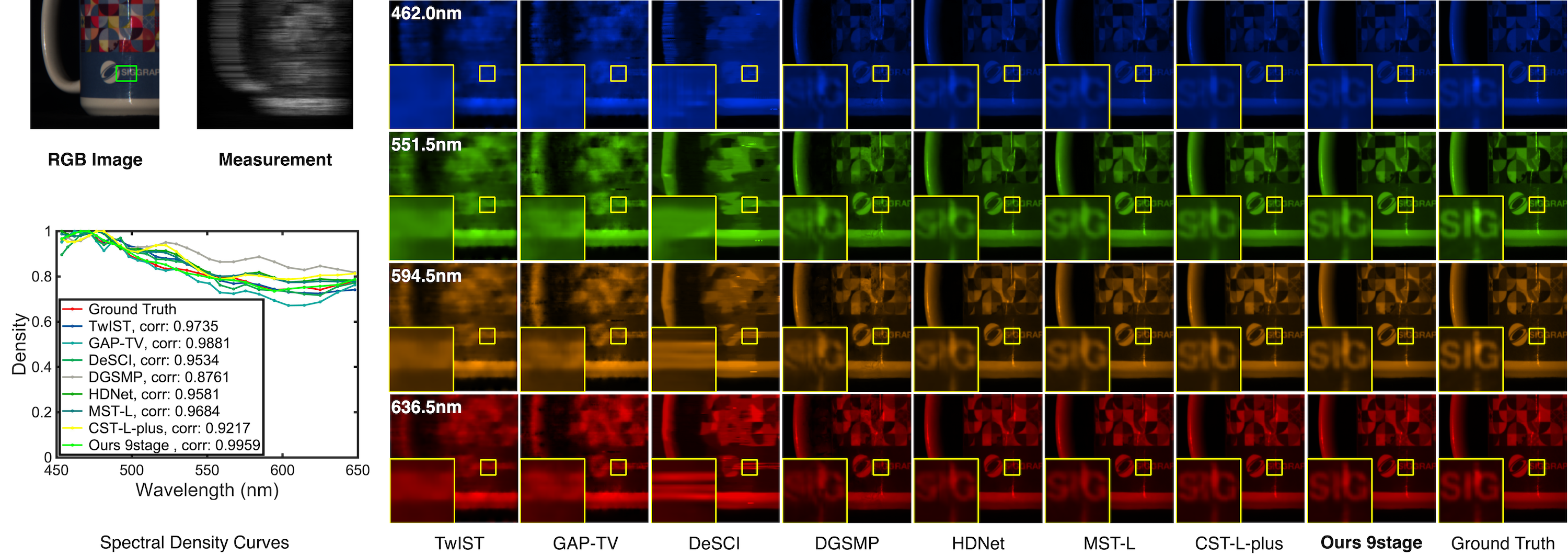}
    \vspace{-5.0mm}
    \caption{\small Comparisons of reconstructed HSIs use 4 out of 28 spectral channels in Scene 5. We evaluated 7 SOTA methods alongside the proposed approach RDLUF-Mix$S^2$ with 9stage. Our method's results are most clear. The region within the green box was chosen for the analysis of the reconstructed spectra.  Zoom in for a more detailed examination.} 
    \label{fig:simu_result}
    \vspace{-5.0mm}
\end{figure*}

\vspace{-2mm}

\subsection{Experimental Settings}

\vspace{-2mm}

We conducted both simulation and real experiments by adopting 28 wavelengths ranging from 450 nm to 650 nm for HSIs. These wavelengths were derived through spectral interpolation manipulation, inspired by the approach used in previous works \cite{meng2020end, huang2021deep, cai2022mask}.

\textbf{Simulation HSI Data.} In our simulation experiment, we used two HSI datasets that are widely used in the field, including CAVE and KAIST \cite{yasuma2010generalized, choi2017high}. The CAVE dataset comprises of 32 HSIs, with a spatial dimension of $512 \times 512$, while the KAIST dataset has 30 HSIs with a spatial dimension of $2704 \times 3376$. Following prior works \cite{meng2020end, huang2021deep, hu2022hdnet, cai2022mask, lin2022coarse}, we used a real mask with a size of $256 \times 256$ during the training process. For the purpose of evaluation, we selected a total of 10 scenes from the KAIST dataset.

\textbf{Real HSI Data.}  In our real experiment, we utilized the HSI dataset acquired through the SD-CASSI system proposed by \cite{meng2020end}. This system captures real-world scenes using 28 wavelengths ranging from 450 to 650 nm and a 54-pixel dispersion. The spatial dimensions of the captured measurements are $660 \times 714$.

\textbf{Evaluation Metrics.} The performance of HSI restoration methods will be assessed through the application of performance measures such as PSNR and SSIM \cite{wang2004image}.

\textbf{Implementation Details.} The RDLUF-Mix$S^2$ model was implemented using the PyTorch framework and trained using the Adam optimizer with hyperparameters $\beta_1 = 0.9$ and $\beta_2 = 0.999$. The training process was performed for a total of 300 epochs using the cosine annealing scheduler with a linear warm-up.  The values of the learning rate and the batch size were configured as $2 \times 10^{-4}$ and 1, respectively. During training, 3D HSI datasets were randomly cropped to generate patches of size $256 \times 256 \times 28$ and $660 \times 660 \times 28$, which were used as labels for the simulation and real experiments respectively. The dispersion shift steps were set to 2. Data augmentation techniques such as random flipping and rotation were employed. The objective of the model was to minimize the Charbonnier loss.

\vspace{-3mm}
\subsection{Quantitative Results}
\vspace{-2mm}

In our study, we performed a comprehensive comparative analysis of the proposed RDLUF-Mix$S^2$ method and SOTA HSI restoration techniques. The techniques included three model-based methods (TwIST \cite{bioucas2007new}, GAP-TV \cite{yuan2016generalized}, and DeSCI \cite{liu2018rank}), as well as seven deep learning-based methods (HSSP \cite{wang2019hyperspectral}, DNU \cite{wang2020dnu}, DGSMP \cite{huang2021deep}, HDNet \cite{hu2022hdnet}, MST \cite{cai2022mask}, CST \cite{lin2022coarse}, and DAUHST \cite{cai2022degradation}). All the techniques were trained using the same datasets and evaluated under the same settings as DGSMP \cite{huang2021deep} to ensure fair comparisons. 
The effectiveness of different methods was evaluated based on the measures of PSNR and SSIM, and the corresponding outcomes for 10 simulated scenes are demonstrated in Table \ref{Tab:performance}. It is noteworthy that while CNN-based and Transformer-based techniques exhibit superior performance than model-based methods, the proposed method surpasses them all. Compared to DGSMP \cite{huang2021deep}, HDNet \cite{hu2022hdnet}, MST-L \cite{cai2022mask}, $\text{CST-L}^*$ \cite{lin2022coarse} and DAUHST-9stg \cite{cai2022degradation}, the proposed method with \emph{9stage} achieves improvements over these methods, which are 6.94 dB, 5.23 dB, 4.39 dB, 3.45 dB and 1.21 dB on average, respectively. Additionally, the proposed method requires cheaper memory costs as shown in Fig. \ref{fig:teaser}.

\begin{figure*}[ht]
    \centering
    \includegraphics[width=\linewidth]{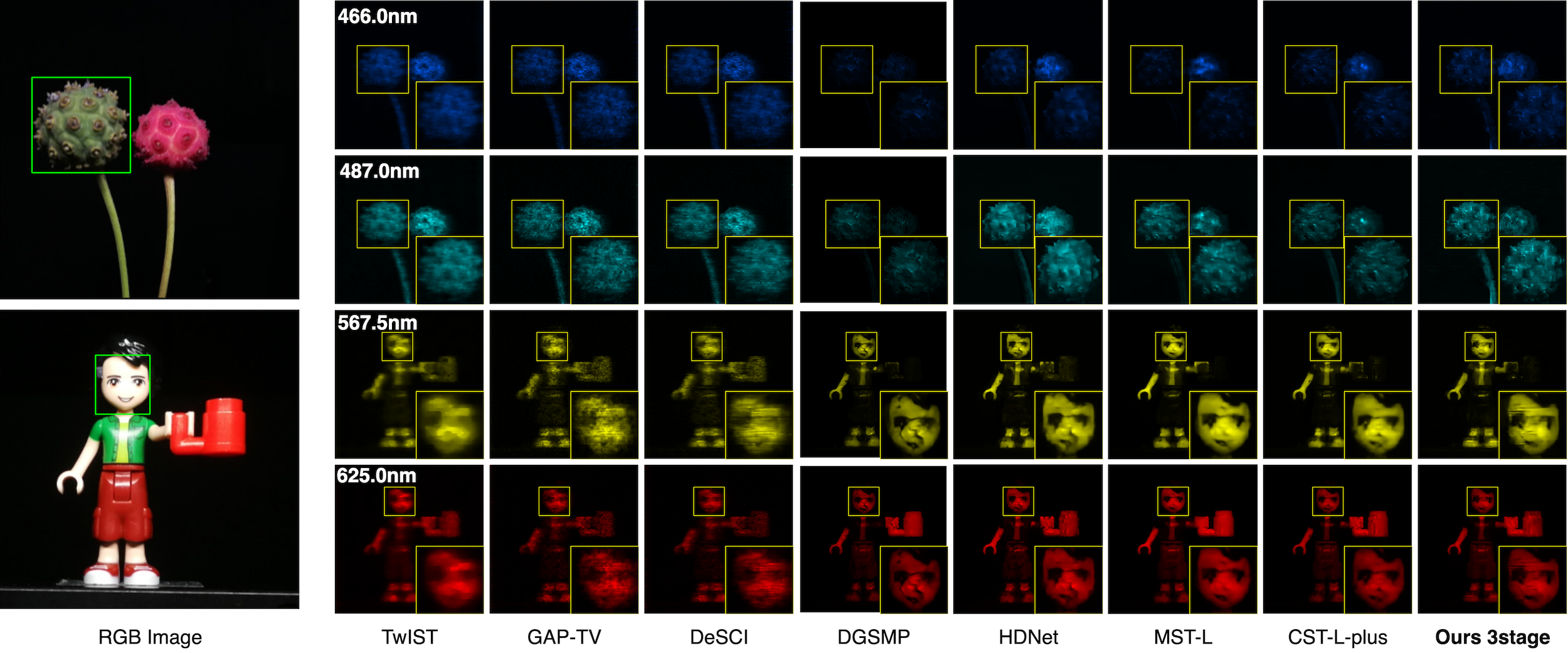}
    \vspace{-2mm}
    \caption{\small Real HSI reconstruction comparison of Scene 3 and Scene 4.  4 out of 28 spectra are randomly selected.}
    \vspace{-3mm}
    \label{fig:real_result}
\end{figure*}

\vspace{-2mm}
\subsection{Qualitative Results}
\vspace{-2mm}

\textbf{Simulation HSI Reconstruction.} We provide a comparison of the proposed RDLUF-Mix$S^2$ method for HSI reconstruction, using 4 of 28 spectral channels of Scene 5, with the simulation results obtained from seven SOTA approaches. As illustrated in Fig. \ref{fig:simu_result}, our method produces visually smoother and cleaner textures, while preserving the spatial information of the homogeneous regions. The results demonstrate that our method is effective in generating high-quality HSIs with improved texture characteristics and spatial information preservation. Specifically, our approach leverages the spectral self-attention branch and the multiscale convolution branch to effectively model long-range dependency and enhance the ability to capture detailed textures respectively. Furthermore, we conducted an evaluation to verify the spectral consistency of our approach by comparing the spectral density curves of the reconstructed areas to the ground truth. As illustrated in the bottom-left of Fig. \ref{fig:simu_result}, our method achieved the highest correlation coefficient, which highlights the effectiveness of our residual degradation learning strategy.

\textbf{Real HSI Reconstruction.} To conduct real experiments, we retrain our model on the CAVE \cite{yasuma2010generalized} and KAIST \cite{choi2017high} datasets and test on real measurements, following the settings of previous works \cite{meng2020end, huang2021deep, cai2022mask, lin2022coarse}. To simulate real measurement conditions, we introduced 11-bit shot noise during the training process. As shown in Fig. \ref{fig:real_result}, we compared the reconstructed images of two real scenes (4 of 28 spectral channels of Scene 3 and Scene 4) using our RDLUF-Mix$S^2$ method and seven SOTA approaches. Our model achieves competitive results with SOTA methods. In Scene 3, the proposed method is able to restore more texture and detail, especially in the cactus areole area. In Scene 4, the proposed method restores the clearer right eye than other methods. Unfortunately, we find that some artifacts were introduced in the real results. This suggests that there are some challenges involved in transferring the trained network from simulated data to real data. As part of future work, we aim to analyze the factors contributing to the artifacts and devise effective measures to mitigate their impact.

\begin{figure}[h]
    \centering
    \includegraphics[width=\linewidth]{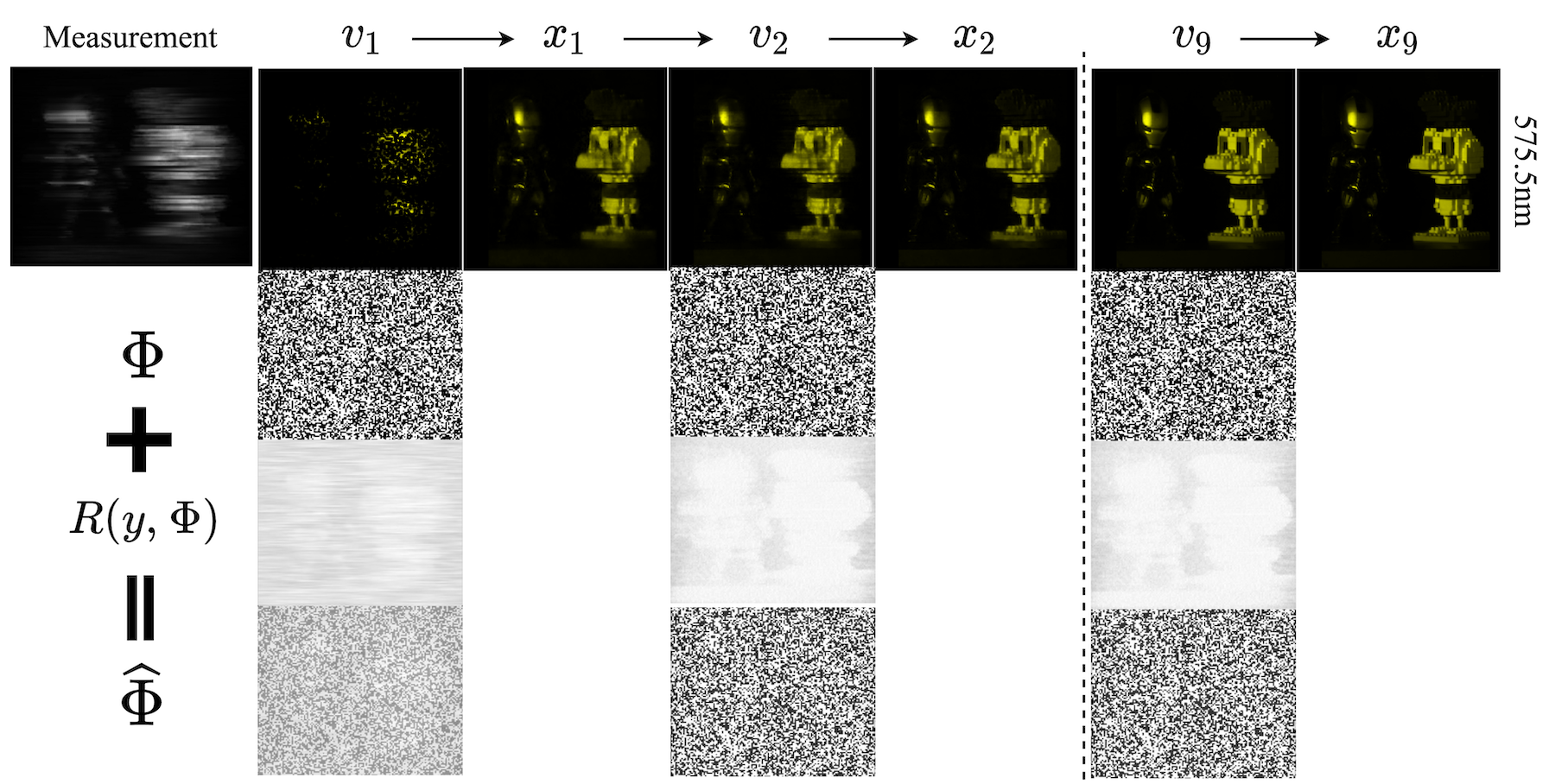}
    \vspace*{-7mm}
    \caption{\small The results of the reconstruction, the visualizations of $\Phi$, the residual $R(y, \Phi)$, and the corrected $\hat \Phi$ are presented. Please note that the visualizations are normalized and the residual response is actually small.} 
    \label{fig:residual}
    \vspace*{-5mm}
\end{figure}

\textbf{Visualization of the Residual Degradation Learning.} We have visually demonstrated the learned degradation matrix and the residual between the sensing matrix and the degradation matrix in Fig. \ref{fig:residual}.  It is observed that the residuals capture useful information such as object contours and minorly correct the sensing matrix. 

\vspace{-2mm}
\subsection{Ablation Study}
\vspace{-2mm}

 \textbf{Break down ablation study.} To investigate the specific impact of the different components of RDLUF-Mix$S^2$ on its overall performance, we conducted an ablation study, and the detailed results are presented in Table \ref{tab:table6}. We first established a baseline model by exclusively employing the spectral branch with a plain gradient descent module, achieving results of 36.49 dB. Subsequently, applying the residual degradation learning strategy resulted in a significant improvement of 0.61 dB. Incorporating both the spectral and spatial branches lead to a further improvement of 0.20 dB. Even with few parameters, the bi-directional interaction managed to boost results by 0.11 dB. Additionally, block interaction and stage interaction yielded improvements of 0.09 dB and 0.06 dB, respectively. By utilizing all components jointly, the method gained a total boost of 1.07 dB, showcasing the effectiveness of RDLUF-Mix$S^2$.

\begin{table}[!htbp]
    \centering
    \scalebox{0.8}{
        \begin{tabular}{c c c  c c c c}
        		%\small
        		\toprule
        		  &  & PSNR & SSIM\\
        		\midrule
        		1 & Baseline(Spectral Branch)  & 36.49 & 0.950 \\
                    2 & 1 + Residual Degradation Learning  & 37.10 \color{green}{\bf (+0.61)} & 0.955 \\
        		  3 & 2 + Spatial Branch & 37.30\color{green}{\bf (+0.20)} & 0.958 \\
                    4 & 3 + Bi-directional Interaction & 37.41\color{green}{\bf (+0.11)} & 0.959 \\
                    5 & 4 + Block Interaction & 37.50\color{green}{\bf (+0.09)} & 0.961 \\
                    6 & 5 + Stage Interaction & \bf 37.56\color{green}{\bf (+0.06)} & \bf{0.963} \\
        		\bottomrule
        \end{tabular}
    }
    \vspace{-0.2cm}
    \caption{\small Break-down ablation studies of every component. }
    \label{tab:table6}
    \vspace{-0.3cm}
\end{table}

\textbf{Number of stages.} Our model shares parameters except for the first and last stages. We investigate the benefits of different numbers of stages, namely 3, 5, 7, and 9 stages. As demonstrated in Tab. \ref{tab:table5}, it can be found that the performance of the network improves with an increase in the number of stages, indicating the efficacy of the iterative network design. Based on a trade-off between reconstruction performance and computational complexity, we find that the 7-stage approach is the optimal choice. Interestingly, we observed that sharing parameters can achieve better performance. It may suggest that a well-trained RDLGD module boosts more than multiple unshared RDLGD modules.

\begin{table}[!htbp]
    \centering
    \scalebox{0.8}{
        \begin{tabular}{c c c  c c c c}
        		%\small
        		\toprule
        		Number of stages  & PSNR & SSIM\\
        		\midrule
        		3  & 37.56 & 0.963 \\
        		5 & 38.59 & 0.969 \\
        		7 & 39.35 & 0.973 \\
        		9 &\bf 39.57 &\bf 0.974 \\
        		$\text{ 9}^*$ & 39.03 & 0.971 \\
        		\bottomrule
        \end{tabular}
    }
    \vspace{-0.2cm}
    \caption{\small Ablation of number of stages. * notes all parameters is independent.}
    \label{tab:table5}
    \vspace{-0.3cm}
\end{table}

%------------------------------------------------------------------------
\vspace{-5mm}
\section{Conclusion and Limitation}
\label{sec:conclusion}
\vspace{-2mm}

In this paper,  we first propose the RDLUF, which bridges the gap between the sensing matrix and the degradation process. Then,  to strengthen the spectral-spatial representation capability in HSI, a Mix$S^2$ Transformer is designed via mixing priors across spectral and spatial. Finally, plugging the Mix$S^2$ Transformer into the RDLUF leads to an end-to-end trainable neural network RDLUF-Mix$S^2$. The proposed method is demonstrated to outperform existing SOTA algorithms on simulation datasets and achieves comparable results to prior works on real datasets.

Although the proposed method achieves a large margin improvement in simulation data, the artifacts in real data suggest that transferring the trained network from simulation to real data could be a challenge. Besides, with the introduction of the convolution branch, more computational complexity is also included.  Therefore, we will analyze the causes of the artifacts and make corrections and exploit some efficient implementations in future research.

\newpage
%%%%%%%%% REFERENCES
{\small
\bibliographystyle{ieee_fullname}
\bibliography{egbib}
}

\end{document}